\documentclass[conference,letterpaper]{IEEEtran}
\IEEEoverridecommandlockouts
\usepackage[utf8]{inputenc} 
\usepackage[T1]{fontenc}
\usepackage{url}
\usepackage{ifthen}
\usepackage{cite}
\usepackage[cmex10]{amsmath} 
\usepackage{amssymb}
\usepackage{multirow}
\usepackage{siunitx}
\usepackage[nolist]{acronym}
\usepackage{tikz}
\usetikzlibrary{decorations.pathreplacing,calligraphy,calc,arrows, arrows.meta, dsp,matrix}
\usepackage{pgfplots}
\usepgfplotslibrary{groupplots}
\usepackage[font=footnotesize]{caption}
\usepackage{subcaption}
\usepackage{bm}
\usepackage{etoolbox} 
\tikzset{>=latex}

\renewcommand{\vec}[1]{\mathbf{#1}}

\newcommand{\cv}{\vec{c}}

\newcommand{\gv}{\vec{g}}
\newcommand{\hv}{\vec{h}}

\newcommand{\mv}{\vec{m}}

\newcommand{\sv}{\vec{s}}

\newcommand{\uv}{\vec{u}}
\newcommand{\vv}{\vec{v}}
\newcommand{\wv}{\vec{w}}
\newcommand{\xv}{\vec{x}}
\newcommand{\yv}{\vec{y}}
\newcommand{\zv}{\vec{z}}

\newcommand{\Hm}{\vec{H}}

\newcommand{\RR}{\mathbb{R}}

\newcommand{\LB}{\left(}
\newcommand{\RB}{\right)}

\renewcommand{\ln}[1]{\mathop{\mathrm{ln}}\LB #1\RB}

\begin{acronym}
 \acro{CSI}{channel state information}
 \acro{NBP}{neural belief propagation}
 \acro{FFG}{Forney factor graph}
 \acro{UFG}{Ungerboeck factor graph}
 \acro{PCM}{parity-check matrix}
 \acro{UE}{user equipment}
 \acro{UL}{uplink}
 \acro{BS}{basestation}
 \acro{TDD}{time division duplex}
 \acro{FDD}{frequency division duplex}
 \acro{ECC}{error-correcting code}
 \acro{MLD}{maximum likelihood decoding}
 \acro{HDD}{hard decision decoding}
 \acro{IF}{intermediate frequency}
 \acro{RF}{radio frequency}
 \acro{SDD}{soft decision decoding}
 \acro{NND}{neural network decoding}
 \acro{CNN}{convolutional neural network}
 \acro{ML}{maximum likelihood}
 \acro{GPU}{graphical processing unit}
 \acro{BP}{belief propagation}
 \acro{LTE}{Long Term Evolution}
 \acro{BER}{bit error rate}
 \acro{DER}{detection error rate}
 \acro{SNR}{signal-to-noise-ratio}
 \acro{ReLU}{rectified linear unit}
 \acro{BPSK}{binary phase shift keying}
 \acro{QPSK}{quadrature phase shift keying}
 \acro{AWGN}{additive white Gaussian noise}
 \acro{MSE}{mean squared error}
 \acro{LLR}{log-likelihood ratio}
 \acro{MAP}{maximum a posteriori}
 \acro{NVE}{normalized validation error}
 \acro{BCE}{binary cross-entropy}
 \acro{CE}{cross-entropy}
 \acro{BLER}{block error rate}
 \acro{SQR}{signal-to-quantisation-noise-ratio}
 \acro{MIMO}{multiple-input multiple-output}
 \acro{OFDM}{orthogonal frequency division multiplex}
 \acro{RF}{radio frequency}
 \acro{LOS}{line of sight}
 \acro{NLoS}{non-line of sight}
 \acro{NMSE}{normalized mean squared error}
 \acro{CFO}{carrier frequency offset}
 \acro{SFO}{sampling frequency offset}
 \acro{IPS}{indoor positioning system}
 \acro{TRIPS}{time-reversal IPS}
 \acro{RSSI}{received signal strength indicator}
 \acro{MIMO}{multiple-input multiple-output}
 \acro{ENoB}{effective number of bits}
 \acro{AGC}{automated gain control}
 \acro{ADC}{analog to digital converter}
 \acro{ADCs}{analog to digital converters}
 \acro{FB}{front bandpass}
 \acro{FPGA}{field programmable gate array}
 \acro{JSDM}{Joint Spatial Division and Multiplexing}
 \acro{NN}{neural network}
 \acro{IF}{intermediate frequency}
 \acro{LoS}{line-of-sight}
 \acro{NLoS}{non-line-of-sight}
 \acro{DSP}{digital signal processing}
 \acro{AFE}{analog front end}
 \acro{SQNR}{signal-to-quantisation-noise-ratio}
 \acro{SINR}{signal-to-interference-noise-ratio}
 \acro{ENoB}{effective number of bits}
 \acro{AGC}{automated gain control}
 \acro{PCB}{printed circuit board}
 \acro{EVM}{error vector mangnitude}
 \acro{CDF}{cumulative distribution function}
 \acro{MRC}{maximum ratio combining}
 \acro{MRP}{maximum ratio precoding}
 \acro{MRT}{maximum ratio transmission}
 \acro{DeepL}{deep-learning}
 \acro{DL}{deep learning}
 \acro{SISO}{single-input single-output}
 \acro{SGD}{stochastic gradient descent}
 \acro{CP}{cyclic prefix}
 \acro{MISO}{Multiple Input Single Output}
 \acro{LMMSE}{linear minimum mean square error}
 \acro{ZF}{zero forcing}
 \acro{USRP}{universal software radio peripheral}
 \acro{RNN}{recurrent neural network}
 \acro{GRU}{gated recurrent unit}
 \acro{LSTM}{long short-term memory}
 \acro{NTM}{neural turing machine}
 \acro{DNC}{differentiable neural computer}
 \acro{TCN}{temporal convolutional network}
 \acro{FCL}{fully connected layer}
 \acro{MLP}{multilayer perceptron}
 \acro{MANN}{memory augmented neural network}
 \acro{RNN}{recurrent neural network}
 \acro{DNN}{deep neural network}
 \acro{FIR}{finite impulse response}
 \acro{BPTT}{back-propagation through time}
 \acro{GAN}{generative adversarial network}
 \acro{ELU}{exponential linear unit}
 \acro{tanh}{hyperbolic tangent}
 \acro{BICM}{bit-interleaved coded modulation}
 \acro{OTA}{over-the-air}
 \acro{IM}{intensity modulation}
 \acro{DD}{direct detection}
 \acro{RL}{reinforcement learning}
 \acro{SDR}{software-defined radio}
 \acro{WGAN}{Wasserstein generative adversarial network}
 \acro{BMD}{bit-metric decoding}
 \acro{BMI}{bit-wise mutual information}
 \acro{LDPC}{low-density parity-check}
 \acro{IDD}{iterative demapping and decoding}
 \acro{IEDD}{iterative equalization, demapping and decoding}
 \acro{JED}{joint equalization and decoding}
 \acro{JSD}{Jensen-Shannon divergence}
 \acro{MMSE}{minimum mean square error}
 \acro{FFT}{fast Fourier transform}
 \acro{IFFT}{inverse fast Fourier transform}
 \acro{QAM}{quadrature amplitude modulation}
 \acro{EMD}{earth mover's distance}
 \acro{TDL}{tapped delay line}
 \acro{KL}{Kullback--Leibler}
 \acro{PRACH}{physical random access channel}
 \acro{URLLC}{ultra-reliable low-latency communication}
 \acro{ANOMA}{asynchronous non-orthogonal multiple access}
 \acro{FEC}{forward error correction}
 \acro{NOMA}{non-orthogonal medium access}
 \acro{MTC}{machine-type communications}
 \acro{mMTC}{massive machine-type communications}
 \acro{MCS}{modulation and coding scheme}
 \acro{PAPR}{peak-to-average power ratio}
 \acro{MAC}{medium access control}
 \acro{STO}{sampling time offset}
 \acro{STE}{straight-through estimator}
 \acro{PHY}{physical}
 \acro{CCE}{categorical cross-entropy}
 \acro{IoT}{Internet of Things}
 \acro{CCDF}{complementary cumulative distribution function}
 \acro{CRC}{cyclic redundancy check}
 \acro{ACLR}{adjacent channel leakage ratio}
\acro{MD}{missed detection}
 \acro{FA}{false alarm}
 \acro{FAR}{false alarm rate}
 \acro{BCJR}{Bahl--Cocke--Jelinek--Raviv}
 \acro{SCNC}{serially concatenated neural code}
 \acro{GNN}{graph neural network}
 \acro{VN}{variable node}
 \acro{CN}{check node}
 \acro{FN}{factor node}
 \acro{DUIDD}{deep-unfolded interleaved detection and decoding}
 \acro{ISI}{inter-symbol interference}
\acro{APP}{a posteriori probability}
\end{acronym}

\definecolor{mittelblau}{RGB}{0, 126, 198}
\definecolor{violettblau}{cmyk}{0.9, 0.6, 0, 0}
\definecolor{rot}{RGB}{238, 28 35}
\definecolor{apfelgruen}{RGB}{140, 198, 62}
\definecolor{gelb}{RGB}{1, 221, 0}
\definecolor{orange}{RGB}{244, 111, 33}
\definecolor{pink}{RGB}{237, 0, 140}
\definecolor{lila}{RGB}{128, 10, 145}
\definecolor{hellgrau}{RGB}{224, 224, 224}
\definecolor{mittelgrau}{RGB}{128, 128, 128}
\definecolor{dunkelgrau}{RGB}{80,80,80}
\definecolor{anthrazit}{RGB}{19, 31, 31}

\pgfplotscreateplotcyclelist{corporate colours markers}{%
rot, every mark/.append style={fill=.!80!rot},mark=*\\%
mittelblau, every mark/.append style={fill=.!80!mittelblau},mark=square*\\%
apfelgruen, every mark/.append style={fill=.!80!apfelgruen},mark=triangle*\\%
orange, mark=star\\%
pink, every mark/.append style={fill=.!80!pink},mark=diamond*\\%
violettblau, every mark/.append style={fill=.!80!violettblau},mark=otimes*\\%
lila, mark=|\\%
gelb, every mark/.append style={fill=.!80!gelb},mark=pentagon*\\%
hellgrau, mark=text,text mark=p\\%
anthrazit, mark=text,text mark=a\\%
}

\pgfplotscreateplotcyclelist{corporate colours markers double}{%
rot, every mark/.append style={fill=.!80!rot},mark=*\\%
rot, dashed, every mark/.append style={fill=.!80!rot,solid},mark=*\\%
mittelblau, every mark/.append style={fill=.!80!mittelblau},mark=square*\\%
mittelblau, dashed, every mark/.append style={fill=.!80!mittelblau,solid},mark=square*\\%
apfelgruen, every mark/.append style={fill=.!80!apfelgruen},mark=triangle*\\%
apfelgruen, dashed, every mark/.append style={fill=.!80!apfelgruen,solid},mark=triangle*\\%
orange, mark=star\\%
orange, dashed, mark=star\\%
pink, every mark/.append style={fill=.!80!pink},mark=diamond*\\%
pink, dashed, every mark/.append style={fill=.!80!pink,solid},mark=diamond*\\%
violettblau, every mark/.append style={fill=.!80!violettblau},mark=otimes*\\%
violettblau, dashed, every mark/.append style={fill=.!80!violettblau,solid},mark=otimes*\\%
lila, mark=|\\%
lila, dashed, mark=|\\%
gelb, every mark/.append style={fill=.!80!gelb},mark=pentagon*\\%
gelb, dashed, every mark/.append style={fill=.!80!gelb,solid},mark=pentagon*\\%
}

\pgfplotsset{
  colormap/magma/.style={%
    /pgfplots/colormap={magma}{%
      rgb=(0.001462, 0.000466, 0.013866)
      rgb=(0.035520, 0.028397, 0.125209)
      rgb=(0.102815, 0.063010, 0.257854)
      rgb=(0.191460, 0.064818, 0.396152)
      rgb=(0.291366, 0.064553, 0.475462)
      rgb=(0.384299, 0.097855, 0.501002)
      rgb=(0.475780, 0.134577, 0.507921)
      rgb=(0.569172, 0.167454, 0.504105)
      rgb=(0.664915, 0.198075, 0.488836)
      rgb=(0.761077, 0.231214, 0.460162)
      rgb=(0.852126, 0.276106, 0.418573)
      rgb=(0.925937, 0.346844, 0.374959)
      rgb=(0.969680, 0.446936, 0.360311)
      rgb=(0.989363, 0.557873, 0.391671)
      rgb=(0.996580, 0.668256, 0.456192)
      rgb=(0.996727, 0.776795, 0.541039)
      rgb=(0.992440, 0.884330, 0.640099)
      rgb=(0.987053, 0.991438, 0.749504)
    },
  },
  colormap/inferno/.style={%
    /pgfplots/colormap={inferno}{%
      rgb=(0.001462, 0.000466, 0.013866)
      rgb=(0.037668, 0.025921, 0.132232)
      rgb=(0.116656, 0.047574, 0.272321)
      rgb=(0.217949, 0.036615, 0.383522)
      rgb=(0.316282, 0.053490, 0.425116)
      rgb=(0.410113, 0.087896, 0.433098)
      rgb=(0.503493, 0.121575, 0.423356)
      rgb=(0.596940, 0.154848, 0.398125)
      rgb=(0.688653, 0.192239, 0.357603)
      rgb=(0.775059, 0.239667, 0.303526)
      rgb=(0.851384, 0.302260, 0.239636)
      rgb=(0.912966, 0.381636, 0.169755)
      rgb=(0.956852, 0.475356, 0.094695)
      rgb=(0.981895, 0.579392, 0.026250)
      rgb=(0.987464, 0.690366, 0.079990)
      rgb=(0.973088, 0.805409, 0.216877)
      rgb=(0.947594, 0.917399, 0.410665)
      rgb=(0.988362, 0.998364, 0.644924)
    },
  },
  colormap/plasma/.style={%
    /pgfplots/colormap={plasma}{%
      rgb=(0.050383, 0.029803, 0.527975)
      rgb=(0.186213, 0.018803, 0.587228)
      rgb=(0.287076, 0.010855, 0.627295)
      rgb=(0.381047, 0.001814, 0.653068)
      rgb=(0.471457, 0.005678, 0.659897)
      rgb=(0.557243, 0.047331, 0.643443)
      rgb=(0.636008, 0.112092, 0.605205)
      rgb=(0.706178, 0.178437, 0.553657)
      rgb=(0.768090, 0.244817, 0.498465)
      rgb=(0.823132, 0.311261, 0.444806)
      rgb=(0.872303, 0.378774, 0.393355)
      rgb=(0.915471, 0.448807, 0.342890)
      rgb=(0.951344, 0.522850, 0.292275)
      rgb=(0.977856, 0.602051, 0.241387)
      rgb=(0.992541, 0.687030, 0.192170)
      rgb=(0.992505, 0.777967, 0.152855)
      rgb=(0.974443, 0.874622, 0.144061)
      rgb=(0.940015, 0.975158, 0.131326)
    },
  },
}

\DeclareRobustCommand{\topbot}{\genfrac{}{}{0pt}{}}

\begin{document}



\title{Graph Neural Network-based\\ Joint Equalization and Decoding } 

\author{%
 \IEEEauthorblockN{Jannis Clausius, Marvin Geiselhart, Daniel Tandler and Stephan ten~Brink}
 \IEEEauthorblockA{Institute of Telecommunications, 
                     Pfaffenwaldring 47, University of Stuttgart, 70569 Stuttgart, Germany\\
                    Email: \{clausius,geiselhart,tandler,tenbrink\}@inue.uni-stuttgart.de}
                    \thanks{This work is supported by the German Federal Ministry of Education and Research (BMBF) within the project Open6GHub (grant no. 16KISK019).}
                    }


\makeatletter
\patchcmd{\@maketitle}  
{\addvspace{0.5\baselineskip}\egroup}
{\addvspace{-1.5\baselineskip}\egroup}
{}
{}
\makeatother

\maketitle
\begin{abstract}
This paper proposes to use \acp{GNN} for equalization, that can also be used to perform \ac{JED}. 
For equalization, the \ac{GNN} is build upon the factor graph representations of the channel, while for \ac{JED}, the factor graph is expanded by the Tanner graph of the \ac{PCM} of the channel code, sharing the \acp{VN}.
A particularly advantageous property of the \ac{GNN} is the robustness against cycles in the factor graphs which is the main problem for \ac{BP}-based equalization.
As a result of having a fully deep learning-based receiver, joint optimization instead of individual optimization of the components is enabled, so-called end-to-end learning.
Furthermore, we propose a parallel flooding schedule that further reduces the latency, which turns out to improve also the error correcting performance. 
The proposed approach is analyzed and compared to state-of-the-art baselines in terms of error correcting capability and latency.
At a fixed low latency, the flooding \ac{GNN} for \ac{JED} demonstrates a gain of 2.25\,dB in \ac{BER} compared to an iterative \ac{BCJR}-BP baseline.
\vspace{-0.15cm}
\end{abstract}

\acresetall
\section{Introduction}
\Ac{BP}-based channel equalization promises lower latency and lower complexity \cite{colavolpe05bpdetection} at the trade-off of sub-optimal performance compared to the \ac{MAP} solution based on the \ac{BCJR} \cite{bahl1974optimal} algorithm. 
Recently, deep learning methods \cite{schmid22neuralBP, liu21nnfactornode, shlezinger2020bcjrnet} were proposed to narrow down the gap between \ac{BP} and \ac{BCJR} for equalization or to improve \ac{BP} performance for channel decoding \cite{nachmani2016learning, cammerer2022gnn}.
These propositions can be categorized as  \ac{DNN}-aided inference oriented by the structure of the underlying factor graph \cite{schlezinger2023model,kschischang01factorgraph}.
In contrast, \cite{nachmani2016learning} and \cite{farsad2018neural} proposed to use general purpose \acp{DNN} (\acp{CNN} and \acp{RNN}, respectively) categorized as model-aided networks.
Here, the path of \ac{DNN}-aided inference is continued as we propose to use a \ac{GNN} for equalization.
The structure and schedule of the \ac{GNN} is the same as in the classical \ac{BP} algorithm.
However, all node and edge calculations are replaced by small \acp{MLP}.
In \cite{cammerer2022gnn} and \cite{satorras2021neural}, \acp{GNN} have been proposed for channel decoding.
In the case of equalization, BCJRNet\cite{shlezinger2020bcjrnet} comes close to our approach, where the \ac{BCJR} algorithm was augmented with small \acp{MLP} as the \acp{FN}.
Also, in \cite{liu21nnfactornode}, a single \ac{MLP}-based \ac{FN} was added to the \ac{BP}-based factor graph for equalization.

Moreover, we propose the use of \acp{GNN} for \ac{JED}, one for equalization and one for decoding. 
Both \acp{GNN} are connected by their \acp{VN} to form a combined factor graph that allows information exchange between equalization and decoding.
In \cite{henkel19jointequalLDPCdecoding}, such a combined graph was used with a decision feedback equalizer and a \ac{BP} decoder.
The joint \ac{GNN} constitutes an fully deep learning-based receiver that can be jointly optimized in an end-to-end fashion.
\acp{DNN} for \ac{JED} have been proposed in \cite{xu18nnjed}, however, the component networks are general purpose \acp{DNN} and not build upon the structure of the underlying factor graph, resulting in limited scalability.
Furthermore, their component networks were jointly optimized, but the \ac{DNN} for equalization did not allow feedback from the decoder. 
Our contributions are as follows:
\begin{itemize}
    \item We demonstrate that \acp{GNN} are capable of closely approaching the \ac{MAP} performance in equalization.
    \item We propose \acp{GNN} for \ac{JED} where the respective factor graphs are connected by their common \acp{VN} and allow joint processing.
    The proposed approach is compared to state-of-the-art baselines.
    \item Finally, we propose a flooding schedule for the \acp{GNN} on the joint factor graph instead of an iterative/sequential schedule, improving both performance and latency.
\end{itemize}







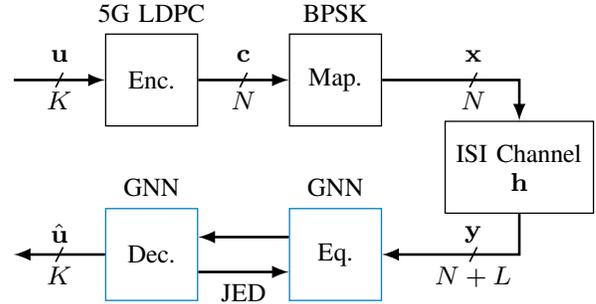
\begin{figure}[tbp]
  \centering
  \resizebox{.89\columnwidth}{!}{\begin{tikzpicture}[
    yscale=0.95,
    font=\footnotesize,
    block/.style={draw, rectangle, minimum size=1cm, align=center},
    gnn/.style={draw, rectangle, minimum size=1cm, align=center, draw=mittelblau,label=GNN},
	mapper/.pic = {
		\begin{axis}[xshift=-0.4cm, yshift=-0.4cm,
			width=0.8cm, height=0.8cm,
			scale only axis, hide axis, clip=false,
			xmin=-1.2, xmax=1.2, ymin=-1.2, ymax=1.2,
			domain=-1:1, samples=100, mark size=1pt]
			\addplot [color=black, mark=*] coordinates {(-0.707,0.707)};
			\addplot [color=black, mark=*] coordinates {(-0.707,-0.707)};
			\addplot [color=black, mark=*] coordinates {(0.707,0.707)};
			\addplot [color=black, mark=*] coordinates {(0.707,-0.707)};
			\addplot[thin] coordinates {(0,-1.2) (0,1.2)};
			\addplot[thin] coordinates {(-1.2,0) (1.2,0)};
		\end{axis}
	},
   tick/.pic = {
    		\draw[line width=0.5pt] (-0.4mm,-0.8mm) -- (0.4mm,0.8mm);
    	}
 ]
    \tikzstyle{arrow} = [thick,->]
\tikzset{
    between/.style args={#1 and #2}{
         at = ($(#1)!0.5!(#2)$)
    }
    }
    \coordinate (source) at (0.5,0);
    \node[block, label = 5G LDPC ](enc) at (2,0) {Enc.} ;
	\node[block, label= BPSK](mapper)at (4,0) {Map.} ; 

    \node[block](ch) at (6,-1) {ISI Channel\\$\mathbf{h}$};
    \node[gnn ](eq) at (4,-2) {Eq.};
    \node[gnn](dec) at (2,-2) {Dec.} ;
	\coordinate (sink) at (0.5,-2);

    \draw[arrow](source) -- node[above,yshift=2pt] {$\uv$} node[] (tickA) {} node[below] {\footnotesize $K$}(enc);
    
    \draw[arrow](enc) -- node[above,yshift=2pt] {$\cv$} node[] (tickB) {} node[below] {\footnotesize $N$}(mapper);
    
    \draw[arrow](mapper) -| node[above,yshift=2pt,xshift = -0.5cm] {$\xv$} node[,xshift = -0.5cm] (tickC) {} node[below, xshift = -0.5cm] {\footnotesize $N$}(ch);

    \draw[arrow]  (ch) |- node[above, xshift=-.5cm] {$\yv$} node[ xshift=-.5cm] (tickD) {} node[below, xshift=-.5cm] {\footnotesize $N+L$}(eq);
    
    \draw[arrow]  ([yshift=-0.2 cm]dec.east) -- node[below, align=center] {\footnotesize JED} ([yshift=-0.2 cm]eq.west);
    
    \draw[arrow]  ([yshift=0.2 cm]eq.west) -- ([yshift=0.2 cm]dec.east);

    \draw[arrow]  (dec) -- node[above] {$\hat{\uv}$} node[] (tickE) {} node[below] {\footnotesize $K$} (sink);
    
    \pic at (tickA) {tick};
    \pic at (tickB) {tick};
    \pic at (tickC) {tick};
    \pic at (tickD) {tick};
    \pic at (tickE) {tick};

\end{tikzpicture}}
  \caption{Block diagram of the system model with an \ac{ISI} channel and \acfp{GNN} for iterative/joint equalization and decoding.}
  \label{fig:systemmodel}
  \vspace{-0.6cm}
\end{figure}
\section{Preliminaries}
\subsection{System Overview}

Fig.~\ref{fig:systemmodel} shows the block diagram of the considered system model. 
We assume 5G \ac{LDPC} channel encoding and \ac{BPSK} at the transmitter.
The \ac{ISI} channel is modeled as a tapped delay line with Proakis-C impulse response ($\hv = [0.227, 0.460, 0.688, 0.460, 0.227]$) \cite{proakis2001digital} and \ac{AWGN}.
Hence, the discrete-time channel can be mathematically described by \cite{forney72observationmodel}
\begin{equation} \label{eq:channel_eq}
    y_k = \sum_{l = 0}^L h_l x_{k-l} + z_k
\end{equation}
with channel memory $L$, transmitted code symbols $x_k$ and where $z_l \sim \mathcal{N} (0,\sigma^2)$ denotes independent and identically distributed  Gaussian noise.
Furthermore, we assume that the symbols $x_k$ for $k \in [ -L+1,-1] \cup [N-1,N+L]$ are known at the receiver, where $N$ denotes the codeword length.
Thus, we can rewrite Eq. (\ref{eq:channel_eq}) to
$ \yv = \mathbf{H} \widetilde{\xv} + \zv $
with Toepelitz matrix $\mathbf{H} \in \RR ^{N+L \times (N+2L)}$, equivalent transmit sequence $\widetilde{\xv} \in \RR^{N+2L}$ and $\zv  \in \RR ^{N+L}$. 
The goal of the receiver is to estimate the bit-wise \acp{APP} $P(u_i|\yv, \Hm)$. 
This is usually done by a sequential processing of equalization and decoding without feedback to the equalizer, which we call \emph{disjoint processing}. 
If the equalizer can refines its estimate based on the decoder feedback, we label it \emph{\acf{JED}}.

\subsection{Belief Propagation}
Consider the factorization of a joint probability function $f(\mathcal{X})$ into a product of $J$ factors $g_j (\mathcal{X}_j)$, i.e., ${f(\mathcal{X}) = \prod_{1 < j \leq J } g_j (\mathcal{X}_j)}$
where $\mathcal{X}$ corresponds to a set of random variables $\{x_1, \dots, x_N\}$ and $\mathcal{X}_j \subset \mathcal{X}$.
One can construct a corresponding factor graph by assigning each variable $x_j$ a \ac{VN} $V_j$, each function $g_i$ a \ac{FN} $F_i$, and introducing an edge connecting between $V_j$ and $F_i$ if $x_j$ is an argument of $g_i$ \cite{kschischang01factorgraph}.
The marginals of $f$, i.e. $f(x_i)$ for each $x_i \in \mathcal{X}$, can be efficiently computed using the message passing algorithm \ac{BP} on the factor graph.
The algorithm operates by sending messages from \acp{VN} (\acp{FN}) to \acp{FN} (\acp{VN}), denoted as $m_{V_i \rightarrow F_j}$ ($m_{F_j \rightarrow V_i} $).
In the log-domain, the messages are determined using 
\begin{align}
   m_{V_i \rightarrow F_j} =& \sum_{F_k \in \mathcal{F}(V_i) \backslash \{F_j\}}  m_{F_k \rightarrow V_i} \\
   m_{F_j \rightarrow V_i}  =& \underset{\sim \{V_i\}}{\operatorname{\max} ^\star} \left( \ln{g_j(\mathcal{X}_j)} + \hspace{-2em} \sum_{ V_k \in \mathcal{V}(F_j) \backslash \{V_i \}} \hspace{-2em} m_{V_k \rightarrow F_j}  \right)
\end{align}
where $\mathcal{V}(F_j)$ and $\mathcal{F}(V_i)$ denote the \emph{neighborhood} of \acp{VN} of $F_j$ and of \acp{FN} of $V_i$, respectively and $\operatorname{\max}^\star_{\sim \{V_i\}}$ denotes the Jacobian algorithm \cite{Robertson95suboptimalMAP} applied over all messages except $m_{V_i \rightarrow F_j}$.
\subsection{Factor Graph-based Equalization}
With an appropriate factorization of the joint symbol \ac{APP} $P(\xv| \mathbf{y,H})$, one can obtain an equivalent factor graph representation.
Here, the \acp{FN} correspond to $ \mathbf{H} \widetilde{\xv}$ and the \acp{VN} to the symbols in $\widetilde{\xv}$ (the first and last $L$ \acp{VN} are \emph{virtual} \acp{VN}).
\Ac{BP} is used to calculate the symbol-wise \ac{APP} $P(x_k|\yv,\mathbf{H})$.
Using Bayes' theorem, we obtain
\begin{equation}
\label{eq:prob}
     P(\xv | \yv) \propto P(\yv | \xv) P(\xv) \propto \exp \left(-\frac{\lVert \yv - \mathbf{H} \widetilde{\xv} \rVert^2 }{2\sigma^2} \right).
\end{equation}
Direct application of the chain rule on Eq. (\ref{eq:prob}) yields
\begin{equation}
\label{eq:forney_factorization}
     P(\yv | \xv) \propto \prod_{k=1}^{N} \exp \left(-\frac{| y_k - \sum_{l = 0}^L h_l \widetilde{x}_{k-l}|^2}{2\sigma^2} \right).
\end{equation}
The factor graph corresponding to the factorization of Eq. (\ref{eq:forney_factorization}) was first proposed in \cite{forney72observationmodel} and is referred to as \ac{FFG} in the following.
By rewriting Eq. (\ref{eq:prob}) and using the substitutions $\mathbf{G} := \mathbf{H}^\mathrm{T} \mathbf{H}$ and $\bm{\chi}:=\mathbf{H}^\mathrm{T} \yv$ we obtain an alternative factorization of the \ac{APP} \cite{colavolpe11ungerboeckdetection, ungerboeck74observationmodel}:
\begin{align}
\label{eq:ungerboeck_factorization}
     P(\yv | \xv) \propto& \prod_{k=1}^{N} \left[ F_k^{\mathrm{UFG}} (x_k) \prod_{\substack{{j=1} ,j \neq k}}^{N} \vspace{-2em}I_{k,j}(x_k,x_j)\right]\\
     F_k^{\mathrm{UFG}} (x_k) :&= \exp \left( \frac{1}{2\sigma^2} (2\chi_k x_k - G_{k,k}|x_k|^2) \right)\\
    I_{k,j} (x_k,x_j) :&= \exp \left( - \frac{1}{2\sigma^2} G_{k,j}x_kx_k \right).
\end{align}
The corresponding factor graph is referred to as \ac{UFG} in the following.
However, applying \ac{BP} to these factor graphs is not guaranteed to converge, since \ac{ISI} channels may contain short cycles in their graphs \cite{colavolpe05bpdetection}.
\begin{figure}[tbp]
  \centering
  \begin{tikzpicture}[
    xscale=1.4,
    squarenode/.style={draw, rectangle, minimum size=1.85em},
    nnnode/.style={draw, rectangle, minimum size=1.2em, inner sep=0pt, fill=white},
    circlenode/.style={draw, circle, minimum size=1.85em,lila},
    arrow/.style={->},
    nnarrow/.style={-{Latex[length=2pt 3 1]}, line width=0.7pt,  mittelblau},
    conn/.style={},
    connd/.style={dashed},
    ]

    \node (l2) at (2,0.8) {\footnotesize $y_{i-2}$};
    \node (l3) at (3,0.8) {\footnotesize $y_{i-1}$};
    \node (l4) at (4,0.8) {\footnotesize $y_{i}$};
    \node (l5) at (5,0.8) {\footnotesize $y_{i+1}$};
    \node (l6) at (6,0.8) {\footnotesize $y_{i+2}$};

    \node[nnnode, draw=white, rotate=-90] (text) at (6.30,0) {\tiny FN};
    \node[nnnode, draw=white, rotate=-90, text=lila] (text) at (6.30,-2) {\tiny VN};
    \foreach \x in {2,...,6} {
        
        \ifthenelse{\x = 4}{\node [squarenode, draw=mittelblau] (c\x) at (\x,0) {\tiny MLP};}{\node [squarenode] (c\x) at (\x,0) {};};
        
        \ifthenelse{\x = 4}{\node [circlenode, draw=mittelblau,text=black] (v\x) at (\x,-2) {\tiny MLP};   }{\node [circlenode] (v\x) at (\x,-2) {};   };
        \ifthenelse{\x = 4}{}{\draw [conn] (c\x) -- (v\x);};
        \draw [arrow] (l\x) -- (c\x);
    }

    \coordinate (v0) at (0,-2);
    \coordinate (v1) at (1,-2);
    \coordinate (c7) at (7,0);
    \coordinate (c8) at (8,0);

    \node at (1.3,0) {$\cdots$};
    \node at (1.3,-2) {$\cdots$};
    \node at (6.7,0) {$\cdots$};
    \node at (6.7,-2) {$\cdots$};

    \node at (2,-2.6) {\footnotesize $i-2$};
    \node at (3,-2.6) {\footnotesize $i-1$};
    \node at (4,-2.6) {\footnotesize $i$};
    \node at (5,-2.6) {\footnotesize $i+1$};
    \node at (6,-2.6) {\footnotesize $i+2$};

    \draw[connd] ( $ (v0)!0.6!(c2) $ ) -- (c2);
    \draw[connd] ( $ (v1)!0.2!(c2) $ ) -- (c2);
    \draw[connd] ( $ (v1)!0.1!(c3) $ ) -- (c3);

    \draw[conn] (v2) -- (c3);
    \draw[conn] (v2) -- (c4);
    \draw[conn] (v3) -- (c4);
    \draw[conn] (v3) -- (c5);
    \draw[conn] (v4) -- (c5);
    \draw[conn] (v4) -- (c6);
    \draw[conn] (v5) -- (c6);

    \draw[connd] (v5) -- ( $ (v5)!0.9!(c7) $ );
    \draw[connd] (v6) -- ( $ (v6)!0.8!(c7) $ );
    \draw[connd] (v6) -- ( $ (v6)!0.4!(c8) $ );

    \draw[nnarrow, mittelblau] ( $ (v2)!0.8!(c4) $ ) -- (c4);
    \draw[nnarrow, mittelblau] ( $ (v3)!0.8!(c4) $ ) -- (c4);

    \draw[nnarrow, mittelblau] ( $ (c6)!0.8!(v4) $ ) -- (v4);
    \draw[nnarrow, mittelblau] ( $ (c5)!0.8!(v4) $ ) -- (v4);

    \node[nnnode, draw=white, rotate=-90] (text) at (4.40,-1) {\tiny MLP};
    \node[nnnode, draw=mittelblau] (unn) at (4.15,-1) {\tikz[baseline=-0.5ex,                                 shorten <=2pt, shorten >=2pt] \draw[-{Latex[length=1.5pt 3 1]}, thin] (0,-.15) -- (0,.15);};

    \node[nnnode, draw=white, rotate=90] (text) at (3.60,-1) {\tiny MLP};
    \node[nnnode, draw=mittelblau] (dnn) at (3.85,-1) {\tikz[baseline=-0.5ex,                                 shorten <=2pt, shorten >=2pt] \draw[-{Latex[length=1.5pt 3 1]}, thin] (0,.15) -- (0,-.15);};

    \draw[nnarrow] (dnn.300) -- (v4.103);
    \draw[nnarrow, mittelblau] (v4.65) -- (unn.255);
    \draw[nnarrow, mittelblau] (v4.130) -- (dnn.263);

    \draw[nnarrow, mittelblau] (unn.123) -- (c4.282);
    \draw[nnarrow, mittelblau] (c4.300) -- (unn.94);
    \draw[nnarrow, mittelblau] (c4.250) -- (dnn.68);
    
    \node[rectangle,dashed, draw=mittelblau, minimum height=3.2cm, minimum width=1.3cm, label=below:\footnotesize GNN structure] (rect) at (4.0,-1.2) {};
\end{tikzpicture}
  \caption{Forney factor graph with \ac{GNN} elements are presented for an exemplary pair of \ac{VN}/\ac{FN} for a channel with memory $L=2$. Blue nodes represent \acp{MLP}.}
  \label{fig:ffg}
    \vspace{-0.5cm}
\end{figure}
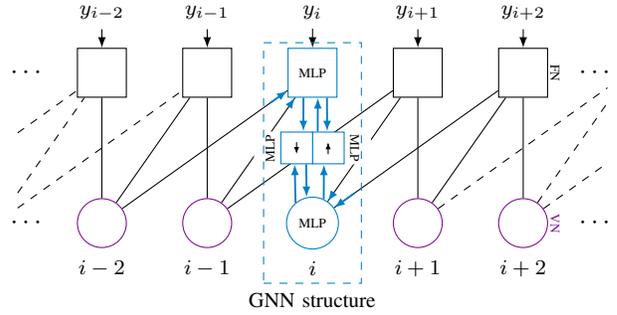


\section{Graph Neural Networks for Equalization}
\label{sec:GNN_eq}

The bipartite \ac{GNN} framework from \cite{cammerer2022gnn} can be directly applied to the \ac{FFG} and \ac{UFG} for equalization. 
In each iteration of the \ac{GNN}, first the \acp{FN} are updated, then the edges towards \acp{VN}, followed by the \acp{VN} themselves, and finally the edges towards \acp{FN}.
As shown in Fig.~\ref{fig:ffg}, the \ac{GNN} assigns learnable parametric functions $f_\theta(\cdot)$ to each \ac{VN}, each \ac{FN} and each directed edge.
Here, this parametric function is implemented by a \ac{MLP} with trainable weights $\theta$.
The weights $\theta$ are shared for each type of node or directed edge in the graph, i.e., there is a set $\theta_V$ for \acp{VN}, $\theta_F$ for \acp{FN}, $\theta_{F\rightarrow V}$ for edges directed towards \acp{VN}, and $\theta_{V\rightarrow F}$ for edges directed towards \acp{FN}.
Each \ac{MLP} represents a node state update function or an edge state update function.
The state of a node or edge is represented by a $d$-dimensional vector, where $d$ is the so called \emph{feature size}.
For simplification, all nodes and edges share the same number of features $d$.
The updated state $\sv^{(t+1)}_{V_i}$ in iteration $t+1$ of node \ac{VN} $V_i$ is calculated by
\begin{equation}
    \sv^{(t+1)}_{V_i} = f_{\theta_{V}} \left( \sv^{(t)}_{V_i}, \frac{1}{|\mathcal{F}(V_i)|}\sum_{F_j \in\mathcal{F}(V_i)} \mv_{F_j \rightarrow V_i} , \gv_{V_i}   \right)
\end{equation}
 where $\mathcal{F}(V_i)$ is the set of connected \acp{FN}, $\mv_{F_j \rightarrow V_i}$ is the state of the edge from $F_j$ to $V_i$, and $\gv_{V_i}\in \RR^d$ is a trainable attribute of node $V_i$.
In a similar fashion, the update of \ac{FN} states $\sv^{(t+1)}_{F_j}$ is calculated by
\begin{equation}
    \sv^{(t+1)}_{F_j} = f_{\theta_{F}} \left( \sv^{(t)}_{F_j} , \frac{1}{|\mathcal{V}(F_j)|}\sum_{V_i \in\mathcal{V}(F_j)} \mv_{V_i \rightarrow F_j} , \gv_{F_j}   \right)
\end{equation}
where where $\mathcal{V}(u_j)$ is the set of connected \acp{VN}.
In between of \ac{VN} and \ac{FN} updates, the edges are updated according to
\begin{align}
     \mv_{V_i \rightarrow F_j} &= f_{\theta_{V\rightarrow F}}\left( \sv_{V_i} ,  \sv_{F_j}, \gv_{V\rightarrow F} \right),   \\ 
    \mv_{F_j \rightarrow V_i} &= f_{\theta_{F\rightarrow V}}\left( \sv_{F_j}, \sv_{V_i} ,\gv_{F\rightarrow V} \right).
\end{align}
While the node states $\sv$ and message states $\mv$ are updated during inference, the attributes $\gv$ are constant during inference, but are learned during training. 
In \cite{cammerer2022gnn}, the node and edge attributes were set to $\mathbf{0}$ as they did not improve the performance for decoding linear channel codes. 
However, in the case of equalization, the edge attributes are crucial, since different edges fulfill different tasks, i.e., in classical \ac{BP} equalization the messages originating from a \ac{FN} are calculated with different functions.
Before and after inference on the graph, the channel outputs are projected to the feature space of the \ac{GNN} and the inference results are projected to \acp{LLR}, respectively.
Before inference, $\yv$ is linearly projected to the $d$ dimensional \ac{FN} state $\sv_{u_i} = \wv y_i$, where $\wv \in \mathbb{R}^{d}$ is a learnable projection vector.
The \ac{VN} states are initialized with $\sv_{v_i} =\mathbf{0}$.
After inference, the \ac{VN} states are projected to \acp{LLR} by $\ell_i = \vv^\mathrm{T}\sv_{v_i}$, where $\vv\in \mathbb{R}^{d}$ denotes a learnable projection vector.
Finally, a sigmoid function $\sigma_{\mathrm{sigmoid}}(\cdot)$ converts the \acp{LLR} to probabilities $\hat{c}_i = \sigma_{\mathrm{sigmoid}}(\ell_i)$.
Note that the conversion from feature vectors to probabilities can be done after any number of iterations, allowing for adjustment to the desired latency, complexity, or performance.

\subsection{Training}
\label{subsec:training_uncoded}
For training, the \ac{BCE} loss $\mathcal{L}_\mathrm{BCE}$ is employed as its minimization was shown to maximize the \ac{BMI}
\begin{equation}
   I_\mathrm{BMI}  = H(\mathsf{U}) - H(\mathsf{U}|\hat{\mathsf{U}}) \overset{\text{\cite{cammerer2019tcom}}}{\approx}    H(\mathsf{U})- \frac{1}{|\mathcal{B}|}\sum_{u,\hat{u}\in \mathcal{B}}\mathcal{L}_\mathrm{BCE}
\end{equation}
where $\mathsf{U}$ and $\hat{\mathsf{U}}$ are the random variables associated with the bits $u$ and bit estimates $\hat{u}$, respectively, and $\mathcal{B}$ is a set of input/output pairs and $|\mathcal{B}|$ is called the batch size. 
While minimizing the \ac{BER} does not directly correspond to maximizing the \ac{BMI}, it was shown in \cite{lian19bploss} that $\mathcal{L}_\mathrm{BCE}$ also works well for minimizing the \ac{BER}.
Furthermore, the \ac{BCE} loss is used in a multi-loss fashion, meaning it is averaged over all $N_\mathrm{It}$ \ac{GNN} iterations $t\in [1,N_\mathrm{It}]$ as
\begin{equation}
        \mathcal{L}_\mathrm{Multi}= \frac{1}{N_\mathrm{It}} \sum_{t=1}^{N_\mathrm{It}}  \mathcal{L}^{(t)} _\mathrm{BCE}.
\end{equation}
Finally, the \ac{BCE} loss is calculated from the coded bits $\cv$ and its estimates $\hat{\cv}$, resulting in
\begin{equation}
        \mathcal{L}^{(t)}_\mathrm{BCE}= \frac{1}{n}\sum_{i=1}^{n} [c_i\log_2\hat{c}_i^{(t)}
        +(1-c_i)\log_2(1-\hat{c}_i^{(t)})].
\end{equation}
This loss was first proposed in \cite{nachmani2016learning} and applied for \ac{GNN}-based decoding in \cite{cammerer2022gnn}.
In addition, we use the Adam \cite{kingma2014adam} optimizer.

\subsection{Results}

\begin{table}[tbp]
  \renewcommand{\arraystretch}{1.3}
  \caption{Hyperparameters of the \acp{GNN} and training}
  \label{tab:table_example}
  \centering
  \setlength\tabcolsep{2pt}
  \begin{tabular}{c l c | l c}
  & NN Parameter & Value & Train. Parameter& Value \\
    \hline
    \hline
    \multirow{4}{*}{\rotatebox{90}{Equalization}} & \# \ac{MLP} hidden layers & 2 & Learning rate & $10^{-4}$ \\
    &\# \ac{MLP} hidden units & 64 & Batch size & $256$ \\
    &activation & ReLU & Epochs Equal. &  $5\cdot10^4$ \\
    & Feature size $d$ & 16 & SNR Equal.&  \qty{10}{dB} to \qty{14}{dB} \\
    \cline{1-5}
    \multirow{2}{*}{\rotatebox{90}{JED\vphantom{q}}}  &Schedule Flooding & ($10$,\,$1$) & Epochs \ac{JED}& $1.6\cdot10^5$ \\
    &Schedule Sequential  & ($3$,\,[$3,5$]) & SNR \ac{JED}&  \qty{10}{dB} to \qty{13}{dB}\\
  \end{tabular}
\end{table}



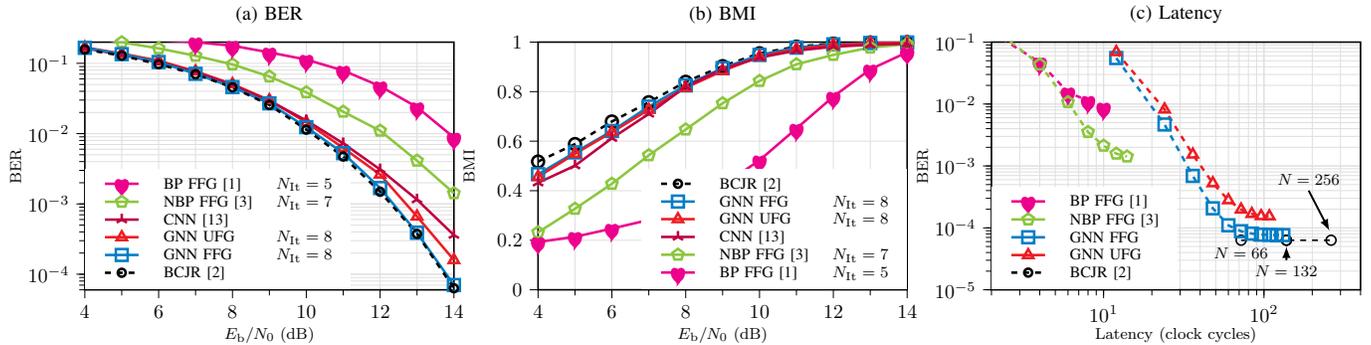
\begin{figure*}[tbp]
  \resizebox{\linewidth}{!}{\begin{tikzpicture}
\begin{groupplot}[
    group style={
        group size= 3 by 1, 
        horizontal sep=1.5cm,
    },
	width=0.45\linewidth,
	height=6cm,
	scale=1,
	minor x tick num=1,
	xmajorgrids,
	xminorgrids=true,
	ymajorgrids,
	yminorticks=true,
    grid=both,
    tick align=outside,
    tick pos=left,
    minor grid style={gray!25},
	major grid style={gray!25},		
    xlabel={\(\displaystyle E_\mathrm{b} \, / \, N_0\) in dB},
    xtick style={color=black},
    ymajorgrids,
    ytick style={color=black},
]
\nextgroupplot[
    title={(a) BER},
    xmode=normal,
    ymode=log,
    xlabel=\footnotesize $E_\mathrm{b}/N_0~(\mathrm{dB})$, %
    ylabel=\footnotesize $\mathrm{BER}$,
    xmin = 4,
    xmax=14,
    ymax=2*10^(-1),
    ymin=6*10^(-5),
    legend style={nodes={scale=0.8, transform shape}},
    legend style={at={(axis cs:8,0.0001)},anchor=south west}, 
    cycle list name=corporate colours markers,
    legend cell align={left},
    line width=0.8pt, %
]
    \addplot+ [rot, mark options={solid},mark=triangle, line width=1.3pt, mark size= 3pt] 
    table[x expr=\thisrowno{0}+0.00 ,y=ber,col sep=comma]{./tikz/results/gnn_ungerboek_proakis_C_n132_it_8.txt};
    \label{plot:gnn_ung_8_ber}

    \addplot+ [apfelgruen, mark options={solid},mark=pentagon, line width=1.3pt, mark size= 3pt] 
    table[x expr=\thisrowno{0}+0.00 ,y=ber,col sep=comma]{./tikz/results/neural_bp_proakis_c_it_7.txt};
    \label{plot:nbp_ffg_7_ber}

    \addplot+ [pink, mark options={solid},mark=heart, line width=1.3pt, mark size= 3pt] 
    table[x expr=\thisrowno{0}+0.00 ,y=ber,col sep=comma]{./tikz/results/proakis_c_FFG_iters_5_n132.txt};
    \label{plot:bp_ffg_5_ber}

    \addplot+ [purple, mark options={solid},mark=Mercedes star, line width=1.3pt, mark size= 3pt] 
    table[x expr=\thisrowno{0}+0.00 ,y=ber,col sep=comma]{./tikz/results/nachmani_cnn.txt};
    \label{plot:nachmani_cnn}

    \addplot+ [mittelblau, mark options={fill=mittelblau, solid},mark=square, line width=1.3pt, mark size= 3pt] 
    table[x expr=\thisrowno{0}+0.00 ,y=ber,col sep=comma]{./tikz/results/gnn_FFG_proakis_C_n132_it_8.txt};
    \label{plot:gnn_ffg_8_ber}

    \addplot+ [black,dashed, mark options={solid},mark=o, line width=1.3pt, mark size= 2pt] 
    table[x expr=\thisrowno{0}+0.00 ,y=ber,col sep=comma]{./tikz/results/proakis_c_BCJR_n132.txt};
    \label{plot:BCJR_ber}

    \coordinate (legend_ber) at (axis description cs:0.0,0.0);

\nextgroupplot[
    title={(b) BMI},
    xmode=normal,
    ymode=normal,
    xlabel=\footnotesize $E_\mathrm{b}/N_0~(\mathrm{dB})$, %
    ylabel=\footnotesize $\mathrm{BMI}$,
    xmin = 4,
    xmax=14,
    ymax=1,
    ymin=0,
    mark size=2.5pt,
    cycle list name=corporate colours markers,
    legend cell align={left},
    line width=0.8pt, %
    ]
    \addplot+ [black,dashed, mark options={solid},mark=o, line width=1.3pt, mark size= 3pt] 
    table[x expr=\thisrowno{0}+0.00 ,y expr=1-(\thisrowno{3}/0.6931),col sep=comma]{./tikz/results/proakis_c_BCJR_n132.txt};
    \label{plot:BCJR_bmi}

    \addplot+ [mittelblau, mark options={fill=mittelblau, solid},mark=square, line width=1.3pt, mark size= 3pt] 
    table[x expr=\thisrowno{0}+0.00 ,y expr=1-(\thisrowno{1}),col sep=comma]{./tikz/results/gnn_FFG_porakis_C_n_132_it_8_bmi_alvarado.txt};
    \label{plot:gnn_ffg_8_bmi}

    \addplot+ [rot,mark options={solid},mark=triangle, line width=1.3pt, mark size= 3pt] 
    table[x expr=\thisrowno{0}+0.00 ,y expr=1-(\thisrowno{1}),col sep=comma]{./tikz/results/gnn_ungerboek_proakis_C_n132_it_8_alvarado.txt};
    \label{plot:gnn_ung_8_bmi}

    \addplot+ [apfelgruen, mark options={solid},mark=pentagon, line width=1.3pt, mark size= 3pt] 
    table[x expr=\thisrowno{0}+0.00 ,y expr=1-(\thisrowno{3}),col sep=comma]{./tikz/results/neural_bp_proakis_c_it_7.txt};
    \label{plot:nbp_ffg_7_bmi}

    \addplot+ [pink, mark options={solid},mark=heart, line width=1.3pt, mark size= 3pt] 
    table[x expr=\thisrowno{0}+0.00 ,y expr=1-(\thisrowno{3}),col sep=comma]{./tikz/results/proakis_c_FFG_iters_5_n132.txt};
    \label{plot:bp_ffg_5_bmi}

    \addplot+ [purple, mark options={solid},mark=Mercedes star, line width=1.3pt, mark size= 3pt] 
    table[x expr=\thisrowno{0}+0.00 ,y expr=1-(\thisrowno{3}),col sep=comma]{./tikz/results/nachmani_cnn.txt};
    \label{plot:nachmani_cnn_bmi}

    \coordinate (legend_bmi) at (axis description cs:1.0,0.0);
    
\nextgroupplot[
    title={(c) Latency},
    xmode=log,
    ymode=log,
    xlabel=\footnotesize $\mathrm{ Latency~(clock~cycles)}$, %
    ylabel=\footnotesize $\mathrm{BER}$,
    xmin = 2,
    xmax=400,
    ymax=10^(-1),
    ymin=10^(-5),
    mark size=2.5pt,
    grid=both,
    cycle list name=corporate colours markers,
    line width=0.8pt, %
]

    \addplot+[mark=o,draw=black, dashed,mark options={solid}]%
        table[x=x,y=y, meta=label] {
        x y label
        72 6.317e-05 c
        138 6.317e-05 c
        138 6.317e-05 c
        262 6.317e-05 c
        };
        \label{plot:latency_bcjr}
 
    \node[] at (axis cs: 72, 0.00004) {\footnotesize $N=66$};
    \node[] at (axis cs: 138, 0.00002) {\footnotesize $N=132$};
    \draw[-latex, line width=1pt, color=black] (axis cs: 138, 3e-5) --node[below,yshift=0cm, xshift=.1cm, color=black]  {} (axis cs: 138, 5e-5);

    \node[] at (axis cs: 190, 0.0006) {\footnotesize $N=256$};
    \draw[-latex, line width=1pt, color=black] (axis cs: 190, 4e-4) --node[below,yshift=0cm, xshift=.1cm, color=black]  {} (axis cs: 256, 9e-5);

    \addplot+ [pink, dashed,mark options={solid},mark=heart, line width=1.3pt, mark size= 3pt] 
    table[x expr=\thisrowno{0}+0.00 ,y=ber,col sep=comma]{./tikz/results/latency/latency_bp_proakis_c_it_5.txt};
    \label{plot:latency_bp}
    
    \addplot+ [apfelgruen, dashed,mark options={solid},mark=pentagon, line width=1.3pt, mark size= 3pt] 
    table[x expr=\thisrowno{0}+0.00 ,y=ber,col sep=comma]{./tikz/results/latency/latency_neural_bp_proakis_c_it_7.txt};
    \label{plot:latency_nbp}

    \addplot+ [mittelblau, dashed,mark options={solid},mark=square, line width=1.3pt, mark size= 3pt] 
    table[x expr=\thisrowno{0}+0.00 ,y=ber,col sep=comma]{./tikz/results/latency/latency_gnn_ffg_uncoded.txt};
    \label{plot:latency_gnn_ffg_uncoded}

    \addplot+ [rot, dashed,mark options={solid},mark=triangle, line width=1.3pt, mark size= 3pt] 
    table[x expr=\thisrowno{0}+0.00 ,y=ber,col sep=comma]{./tikz/results/latency/latency_gnn_ung_uncoded.txt};
    \label{plot:latency_gnn_ung_uncoded}
    
    \coordinate (legend_latency) at (axis description cs:0.0,0.0);

\end{groupplot}		

\node (tab) [
    shape=rectangle,
    draw=none,
    anchor=south west,
    fill=white,
    fill opacity=0.7,
    text opacity=1,
    font=\footnotesize
] at (legend_ber){
\begin{tabular}{clc}
    \ref{plot:bp_ffg_5_ber} & BP \ac{FFG} \cite{colavolpe05bpdetection}& $N_\mathrm{It}=5$ \\
	\ref{plot:nbp_ffg_7_ber} & NBP \ac{FFG} \cite{schmid22neuralBP}& $N_\mathrm{It}=7$ \\
     \ref{plot:nachmani_cnn} & CNN \cite{xu18nnjed} &  \\
    \ref{plot:gnn_ung_8_ber} & GNN \ac{UFG} & $N_\mathrm{It}=8$ \\
	\ref{plot:gnn_ffg_8_ber} & GNN \ac{FFG}  & $N_\mathrm{It}=8$ \\
	\ref{plot:BCJR_ber} & BCJR \cite{bahl1974optimal} &  \\
\end{tabular}
};

\node (tab) [
    shape=rectangle,
    draw=none,
    anchor=south east,
    fill=white,
    fill opacity=0.9,
    text opacity=1,
    font=\footnotesize
] at (legend_bmi){
\begin{tabular}{clc}
    \ref{plot:BCJR_ber} & BCJR \cite{bahl1974optimal} &  \\
    \ref{plot:gnn_ffg_8_ber} & GNN \ac{FFG}  & $N_\mathrm{It}=8$ \\
     \ref{plot:gnn_ung_8_ber} & GNN \ac{UFG} & $N_\mathrm{It}=8$ \\
    \ref{plot:nachmani_cnn} & CNN \cite{xu18nnjed} &  \\
    \ref{plot:nbp_ffg_7_ber} & NBP \ac{FFG} \cite{schmid22neuralBP}& $N_\mathrm{It}=7$ \\
    \ref{plot:bp_ffg_5_ber} & BP \ac{FFG} \cite{colavolpe05bpdetection}& $N_\mathrm{It}=5$ \\

\end{tabular}
};

\node (tab) [
    shape=rectangle,
    draw=none,
    anchor=south west,
    fill=white,
    fill opacity=0.7,
    text opacity=1,
    font=\footnotesize
] at (legend_latency){
\begin{tabular}{cl}
    \ref{plot:bp_ffg_5_ber} & BP \ac{FFG} \cite{colavolpe05bpdetection} \\
	\ref{plot:nbp_ffg_7_ber} & NBP \ac{FFG} \cite{schmid22neuralBP} \\
 \ref{plot:gnn_ffg_8_ber} & GNN \ac{FFG}  \\
    \ref{plot:gnn_ung_8_ber} & GNN \ac{UFG} \\
	\ref{plot:BCJR_ber} & BCJR \cite{bahl1974optimal}   \\
\end{tabular}
};

\end{tikzpicture}}
  \caption{\footnotesize Uncoded BER, BMI and latency of different equalizers for the Proakis-C channel and block length $N=132$. For the BER and BMI plot, $N_\mathrm{It}$ is selected such that increasing it does not improve the performance. Latency is given at $E_\mathrm{b}/N_0 = \qty{14}{dB}$, and graphed over different values of $N_\mathrm{It}$. Note that the latency of the \ac{BCJR} is the only value depending on the sequence length due to its sequential nature. }
  \label{fig:all_uncoded}
   \vspace{-0.4cm}
\end{figure*}

For the evaluation of the proposed \ac{GNN} equalizer, a  transmission of $N=132$ \ac{BPSK} symbols over the Proakis-C channel is simulated. The channel is chosen to represent a scenario with severe \ac{ISI}, where \ac{BPSK} signaling results in at least an estimated loss of \qty{2.01}{dB} and \qty{5.01}{dB} compared to an \ac{AWGN} channel, for \ac{JED} and disjoint processing, respectively \cite{roumy99proakiscloss}. 
We assume the channel to be fixed throughout the paper.
Therefore, the results demonstrate the algorithmic capability of \ac{GNN}-based equalization rather than generalization, robustness and adaptability aspects.
The hyperparameters for the structure of the \ac{GNN} and the training are shown in the top part of Tab.~\ref{tab:table_example}. 
Fig.~\ref{fig:all_uncoded} (a) and (b) compare the \ac{BER} and \ac{BMI} of the \ac{GNN} equalizer based on the \ac{FFG} and \ac{UFG}.
Note that for Fig.~\ref{fig:all_uncoded} (b) all outputs are multiplied with a damping factor to increase the \ac{BMI} \cite{szczecinski2015bit}.
The \ac{BCJR}-based equalizer implements a \ac{MAP} estimator and serves as a baseline.
Furthermore, we show a \ac{CNN} for equalization with the same structure as in \cite{xu18nnjed} but with an increased number of feature maps (to $100$) to achieve a competitive performance for this channel.
The other baselines are, similar to the \ac{GNN}, based on the \ac{BP} algorithm on the \ac{FFG}:
\ac{BP} without neural augmentations and \ac{NBP} with learned multiplicative weights per edge per iteration.
Note that the \ac{BP} equalizer diverges after $N_\mathrm{It}=5$ iterations due to cycles in the graph.
Training of the \ac{NBP} does not converge for $N_\mathrm{It}>7$.
The \acp{GNN} show substantial gains, with the \ac{FFG} variant outperforming the \ac{UFG}, suggesting strong robustness to the cycles in the graph.
First, this matches the results for \ac{NBP} on the \ac{FFG} compared to the \ac{UFG} \cite{schmid22neuralBP}.
Second, the \ac{FFG}-based \ac{GNN} is significantly less complex than the \ac{UFG} variant because the complexity scales with the number of nodes and edges in the graph rather than the \ac{FN} degree as for classical \ac{BP} variants \cite{colavolpe05bpdetection}.
The accumulated number of nodes and edges in the \ac{FFG} is $N(L+3)+L(L+4) \approx N(L+3)$, whereas for the \ac{UFG} it is $N(3L+3) +L(3L+4)\approx N(3L+3)$.
For both factor graphs, we prune the first and last $L$ virtual \acp{VN} without any loss in performance.
An interesting observation is that the performance of the \ac{FFG}-\ac{GNN} remains close to \ac{MAP} even in the high \ac{SNR} regime, while \ac{NBP} variants tend to diverge \cite{schmid22neuralBP}.
Finally, Fig.~\ref{fig:all_uncoded} (c) shows the \ac{BER} vs. latency in clock cycles.
We assume that every node or neural network layer is processed in one clock cycle.
This means $12$ and $2$ cycles per iteration for the \acp{GNN} and (N)BP, respectively.
The latency of the \ac{BCJR} equalizer equates to $N+L+2$, and is, thus, dependent on the block length.
We assume a parallel computation of the forward and backward metric in $N+L$ cycles and $2$ additional cycles are needed to calculate the channel likelihoods and the bit estimates. 
As a result, the \ac{GNN} shows larger latency gains as the transmission length increases.

\section{Graph Neural Networks for Joint Equalization and Decoding}

\label{sec:jed}
In this section, we describe a \ac{DNN} for \ac{JED} based on \acp{GNN}.
A fully deep learning-based receiver allows to jointly optimize the component networks rather than an individual optimization like in traditional receivers.
This is called end-to-end learning. 
Furthermore, using \acp{GNN} for the components prevents the \acp{NN} from the \emph{curse of dimensionality} (referring to exponential growth of points in high-dimensional objects) \cite{bellman61curseofdimensionality}, since the graph-based structures leverage the sparseness of the connections.
Moreover, the \acp{GNN} use high-dimensional processing the nodes and edges, leveraging the \emph{blessing of dimensionality} \cite{donoho00blessingofdimensionality}.

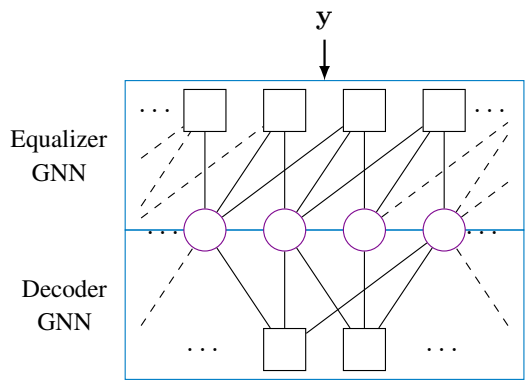
\begin{figure}[tbp]
  \centering
  \resizebox{.8\columnwidth}{!}{\begin{tikzpicture}[
    yscale=.75,
    rotate=90,
    block/.style={draw, rectangle, minimum size=1cm, align=center},
    squarenode/.style={draw, rectangle, minimum size=1.5em},
    circlenode/.style={draw, circle, minimum size=1.5em, lila, fill=white},
    arrow/.style={->,thick},
    noarrow/.style={thick},
    conn/.style={},
    connd/.style={dashed},
    ]

\draw[mittelblau] (-.5,-2.5) rectangle (2,2.5);
\draw[mittelblau] (2,-2.5) rectangle (4.5,2.5);

    \node [squarenode] (dc1) at (0,.5) {};
    \node [squarenode] (dc2) at (0,-.5) {};
    \coordinate (dc3) at (0,-2.5);
    \coordinate (dc4) at (0,2.5);
    
    \node [circlenode] (v1) at (2,1.5) {};
    \node [circlenode] (v2) at (2,.5) {};
    \node [circlenode] (v3) at (2,-.5) {};
    \node [circlenode] (v4) at (2,-1.5) {};
    \draw[conn] (dc1) -- (v1);
	\draw[conn] (dc1) -- (v2);
	\draw[conn] (dc1) -- (v4);
	\draw[conn] (dc2) -- (v2);
    \draw[conn] (dc2) -- (v3);
	\draw[conn] (dc2) -- (v4);

 \draw[connd] ( $ (dc3)!0.2!(v4) $ ) -- (v4);
 \draw[connd] ( $ (dc4)!0.2!(v1) $ ) -- (v1);

    \coordinate (vm1) at (2,3.5);
    \coordinate (v0) at (2,2.5);

    \node at (4,2.1) {$\dots$};
    \node at (4,-2.1) {$\dots$};
    \node at (0,-1.5) {$\dots$};
    \node at (0,1.5) {$\dots$};
    \node at (1.95,2.0) {$\dots$};
    \node at (1.95,-2.0) {$\dots$};
    
    \node [squarenode] (ec1) at (4,1.5) {};
    \node [squarenode] (ec2) at (4,0.5) {};
    \node [squarenode] (ec3) at (4,-0.5) {};
    \node [squarenode] (ec4) at (4,-1.5) {};
    \coordinate (ec5) at (4,-2.5);
    \coordinate (ec6) at (4,-3.5);
    
    \draw[conn] (v1) -- (ec1);
	\draw[conn] (v1) -- (ec2);
	\draw[conn] (v1) -- (ec3);
	\draw[conn] (v2) -- (ec2);
    \draw[conn] (v2) -- (ec3);
    \draw[conn] (v2) -- (ec4);
    \draw[conn] (v3) -- (ec3);
    \draw[conn] (v3) -- (ec4);
    \draw[conn] (v4) -- (ec4);
    
    \draw[connd] ( $ (vm1)!0.6!(ec1) $ ) -- (ec1);
    \draw[connd] ( $ (v0)!0.2!(ec1) $ ) -- (ec1);
    \draw[connd] ( $ (v0)!0.1!(ec2) $ ) -- (ec2);

    \draw[connd] (v3) -- ( $ (v3)!0.9!(ec5) $ );
    \draw[connd] (v4) -- ( $ (v4)!0.8!(ec5) $ );
    \draw[connd] (v4) -- ( $ (v4)!0.4!(ec6) $ );

    \coordinate (input) at (4.5,0);
    \node (y) at (5.5,0) {$\yv$};

    \draw[arrow] (y) -- (input);

    \node[anchor=east,align=center] at (0.75,2.6) {\small Decoder\\\small GNN};
    \node[anchor=east,align=center] at (3.25,2.6) {\small Equalizer\\\small GNN};
	
\end{tikzpicture}}
  \caption{\ac{JED} system with a decoder \ac{GNN} based on the Tanner graph of the channel code and an equalizer \ac{GNN} based on the \ac{FFG} of the channel. Note that the Tanner graph of the channel code code has been rearranged according to the interleaver to directly match the equalizer \acp{VN}.}
  \label{fig:combined_graph}
  \vspace{-0.6cm}
\end{figure}

The structure of the \ac{GNN} for \ac{JED} is displayed in Fig.~\ref{fig:combined_graph}.
One set of \acp{FN} is based on the graph for equalization as in Sec.~\ref{sec:GNN_eq} and one of \acp{FN} for decoding as in \cite{cammerer2022gnn}. 
Note, both graphs share the same \acp{VN}, building a connected graph. 
Furthermore, interleaving and deinterleaving makes the a priori information of the equalizer and decoder to appear more independent, thus, improving performance.


\subsection{Schedule}
\begin{figure}[tbp]
  \centering
       \begin{subfigure}[b]{0.48\columnwidth}
       \centering
            \begin{tikzpicture}[
    yscale=0.7,
    dot/.style={circle, draw=none,fill=black,inner sep=1pt},
    syncnode/.style={circle, draw=none,fill=lila,inner sep=1pt},
    arrow/.style={->},
    ]
    
    \node at (2,-.5) {\footnotesize Eq. FN\vphantom{q}};
    \node[lila] at (1,-.5) {\footnotesize VN\vphantom{q}};
    \node at (0,-.5) {\footnotesize Dec. FN\vphantom{q}};

    \node at (-.8,-.5) {\footnotesize It.\vphantom{q}};
    \foreach \x in {1,...,4} {
        \node at (-.8,-\x) {\footnotesize \x};
        \node [dot] (d\x) at (0,-\x) {};
        \node [dot] (e\x) at (2,-\x) {};
    }
    \foreach \x in {1,...,3} {
        \node [syncnode] (s\x) at (1,-\x-.5) {};
        \draw[arrow] (e\x) -- (s\x);
        \draw[arrow] (d\x) -- (s\x);
        \pgfmathtruncatemacro{\xp}{\x+1}
        \draw[arrow] (s\x) -- (e\xp);
        \draw[arrow] (s\x) -- (d\xp);

        \draw[arrow, densely dashed] (d\x) to[out=210,in=150]  (d\xp); 
        \draw[arrow, densely dashed] (e\x) to[out=330,in=30]  (e\xp);
    }
    \node[align=center, rotate=270] at (2.5,-1.5) {  \scriptsize FN state};
\end{tikzpicture}
            \caption{ Flooding }
            \label{subfig:flooding}
        \end{subfigure}
        \hfill
        \begin{subfigure}[b]{0.48\columnwidth}
            \centering
            \begin{tikzpicture}[
    yscale=0.7,
    dot/.style={circle, draw=none,fill=black,inner sep=1pt},
    syncnode/.style={circle, draw=none,fill=lila,inner sep=1pt},
    arrow/.style={->},
    ]
    
    \node at (2,-.5) {\footnotesize Eq. FN\vphantom{q}};
    \node[lila] at (1,-.5) {\footnotesize VN\vphantom{q}};
    \node at (0,-.5) {\footnotesize Dec. FN\vphantom{q}};

    \node at (-.8,-.5) {\footnotesize It.\vphantom{q}};
    \foreach \x in {1,...,4} {
        \node at (-.8,-\x) {\footnotesize \x};
        \node [dot] (d\x) at (0,-\x) {};
        \node [dot] (e\x) at (2,-\x) {};
    }
    \foreach \x in {1,...,3} {
        \node [syncnode] (s\x) at (1,-\x-.5) {};
    }   
    \draw[arrow] (e1) -- (s1);
    \draw[arrow] (s1) -- (d2);
    \draw[arrow] (d2) -- (s2);
    \draw[arrow] (s2) -- (e3);
    \draw[arrow] (e3) -- (s3);
    \draw[arrow] (s3) -- (d4);    
    
    \draw[arrow, dashed] (d2) to[out=210,in=150] (d4); 
    \draw[arrow, dashed] (e1) to[out=330,in=30]  (e3);
    
\end{tikzpicture}
            \caption{ Sequential }
            \label{subfig:sequential}
        \end{subfigure}
  \caption{\footnotesize Schedules of the \ac{GNN}-based \ac{JED}}
  \label{fig:schedule}
   \vspace{-0.3cm}
\end{figure}
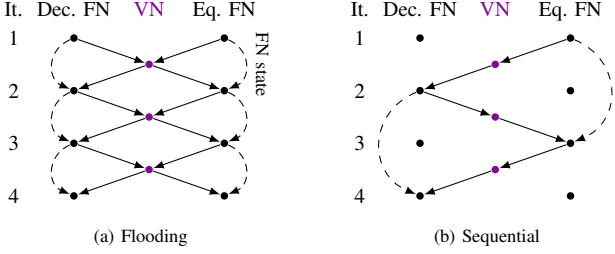

In contrast to the classical solution of using iterative \ac{BCJR} equalization and \ac{BP} decoding \cite{douillard95turboequalization}, the combined \ac{GNN} for joint equalization and decoding enables an earlier use of equalization with a prioi knowledge.
\ac{BCJR} runs are costly in complexity and latency, thus, a sufficient amount of iterations should be spend on \ac{BP} decoding to generate a significant amount of a priori information for the subsequent \ac{BCJR} run.
However, during this time, \ac{BP} decoding does not benefit from new a priori information from the equalizer. 
In case of the combined \ac{GNN}, the latency for equalization and decoding are the same, thus, enabling schedules with earlier information exchange between the components.
This can result in improved error correcting performance at the same latency. 
Moreover, the combined \ac{GNN} can be updated in a flooding fashion. 
The difference to the iterative, sequential schedule is shown in Fig.~\ref{fig:schedule}.
The \ac{GNN} updates the \acp{FN} for equalization and for decoding simultaneously and combines their beliefs after each iteration in their shared \acp{VN}.
Consequently, the same number of resources can be used as in the iterative case, but in half the time.
In every iteration, the component \acp{GNN} receive new a priori information but also keep their respective \ac{FN} state.
In \cite{wiesmayr2022duidd}, it was found that keeping the state of the \acp{FN} is beneficial for iterative systems with a small number of inner iterations.

\subsection{Training}
The training is similar to the training process described in section \ref{subsec:training_uncoded} with some extensions.
Note that the 5G \ac{LDPC} code contains punctured \acp{VN} that we include in the calculation of the loss.
The additional hyperparamters are shown below the horizontal line in Tab.~\ref{tab:table_example}.
The schedule of the combined \acp{GNN} is given in the form ($\#\mathrm{Outer~Iterations}$, [$\#\mathrm{Inner~Iterations}$]).
For deep \acp{GNN}, i.e., a large number of iterations, direct training led to sub-optimal performance. 
Therefore, we used a two stage training process in these cases.
First, we trained with the schedule given in Tab.~\ref{tab:table_example}, for which the \acp{GNN} trained reliable and consistent.
Second, the number of iterations is increased to the desired schedule and the model is finetuned. 
A flexible schedule is ensured by the weight-sharing over iterations in the \acp{GNN}.


\subsection{Results}

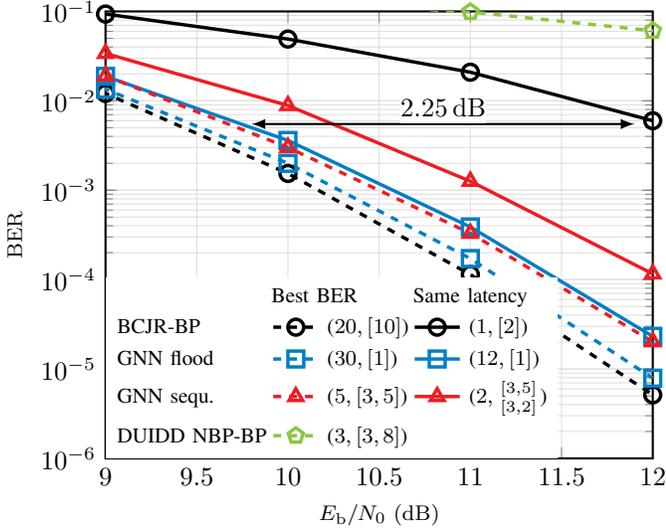
\begin{figure}[t]
	\centering
	\begin{tikzpicture}
		
		\pgfplotsset{compat=1.10}
		
		\begin{axis}[
			xmode=normal,
			ymode=log,
			xlabel=\footnotesize $E_\mathrm{b}/N_0~(\mathrm{dB})$, %
			ylabel=\footnotesize $\mathrm{BER}$,
			xmin = 9,
			xmax=12,
			ymax=1*10^(-1),
			ymin=10^(-6),
			mark size=2.5pt,
			legend style={at={(axis cs:8,0.1)},anchor=south west}, 
			grid=both,
			minor grid style={gray!25},
			major grid style={gray!25},
			width=\linewidth,
	        height=0.85\linewidth,
			cycle list name=corporate colours markers,
			legend cell align={left},
			line width=0.8pt, %
			]

            \addplot+ [black, dashed,mark options={solid},mark=o, line width=1.3pt, mark size= 3pt] 
			table[x expr=\thisrowno{0}+0.00 ,y=ber,col sep=comma]{./tikz/results/coded/coded_proakis_C_BCJR_5GLDPC_k66_n132_SPA_10_IDD_20.txt};
			\label{plot:coded_proakis_C_BCJR_5GLDPC_k66_n132_SPA_20_IDD_10_supplement_best}

   \addplot+ [black, mark options={solid},mark=o, line width=1.3pt, mark size= 3pt] 
			table[x expr=\thisrowno{0}+0.00 ,y=ber,col sep=comma]{./tikz/results/coded/coded_proakis_C_BCJR_5GLDPC_k66_n132_SPA_2_IDD_0.txt};

			\label{plot:coded_proakis_C_BCJR_5GLDPC_k66_n132_SPA_20_IDD_10_supplement}

            \addplot+ [mittelblau, mark options={fill=mittelblau, solid},mark=square, line width=1.3pt, mark size= 3pt] 
			table[x expr=\thisrowno{0}+0.00 ,y=ber_12,col sep=comma]{./tikz/results/coded/gnn_JED_Flood_5G_proakis_C_k66_n132_supp.txt};
			\label{plot:gnn_flood}
            
            \addplot+ [mittelblau, dashed,mark options={fill=mittelblau, solid},mark=square, line width=1.3pt, mark size= 3pt] 
			table[x expr=\thisrowno{0}+0.00 ,y=ber,col sep=comma]{./tikz/results/coded/gnn_JED_Flood_5G_proakis_C_k66_n132.txt};
			\label{plot:gnn_flood_best}

            \addplot+ [rot,dashed, mark options={ solid},mark=triangle, line width=1.3pt, mark size= 3pt] 
			table[x expr=\thisrowno{0}+0.00 ,y=ber,col sep=comma]{./tikz/results/coded/gnn_JED_Seq_5G_proakisC_k66_n132_sched_35_35_35_finetune_35_35_35_35_35.txt};
			\label{plot:gnn_seq_best}

            \addplot+ [rot, mark options={ solid},mark=triangle, line width=1.3pt, mark size= 3pt] 
			table[x expr=\thisrowno{0}+0.00 ,y=ber_7,col sep=comma]{./tikz/results/coded/gnn_JED_Seq_5G_proakisC_k66_n132_sched_35_35_35_results_supplement.txt};
			\label{plot:gnn_seq}

            \addplot+ [apfelgruen, dashed,mark options={ solid},mark=pentagon, line width=1.3pt, mark size= 3pt] 
			table[x expr=\thisrowno{0}+0.00 ,y=ber,col sep=comma]{./tikz/results/coded/duidd_3_bp_nn_3_sp_8.txt};
			\label{plot:duidd_best}

        \coordinate (legend) at (axis description cs:0.0,0.0);

        \draw[<->] (axis cs: 11.9,5.5e-3) -- node[above, rectangle, minimum size=1.2em, inner sep=0pt, fill=white]{$2.25\operatorname{dB}$} (axis cs: 9.8,5.5e-3);

		\end{axis}

  \matrix [
draw=none,
fill=white,
draw opacity = 0.5,
fill opacity=0.7,
text opacity=1,
opacity=0.1,
matrix of nodes,
align =left,
column sep = -5.,
inner sep= 2,
anchor=south west,
font=\footnotesize,
column 1/.style={anchor=base west},
column 2/.style={anchor=base west},
column 3/.style={anchor=base west},
mark options={solid},
] at (legend) {
	& Best BER & Same latency \\
	BCJR-BP & \ref{plot:coded_proakis_C_BCJR_5GLDPC_k66_n132_SPA_20_IDD_10_supplement_best} ($20,[10]$)& \ref{plot:coded_proakis_C_BCJR_5GLDPC_k66_n132_SPA_20_IDD_10_supplement} ($1,[2]$)\\
    GNN flood & \ref{plot:gnn_flood_best} ($30,[1]$) & \ref{plot:gnn_flood} ($12,[1]$)\\
    GNN sequ. & \ref{plot:gnn_seq_best} ($5,[3,5]$) & \ref{plot:gnn_seq} ($2,\topbot{[3,5]}{[3,2]}$)\\
    DUIDD NBP-BP & \ref{plot:duidd_best} ($3,[3,8]$)  &  \\
};

	\end{tikzpicture}
  \vspace{-0.6cm}
	\caption{\footnotesize \ac{BER} of uncoded bits $\uv$ using coded system ($N=132,~R=0.5$, $5$G \ac{LDPC},  \ac{BPSK}) over \ac{SNR} of different equalizers and decoders over the Proakis-C channel. Curves with \emph{best BER} were the best achieved results disregarding latency or complexity. Curves with \emph{same latency} use $144$ clock cycles ($12$ \ac{GNN} flooding iterations). The \ac{JED} schedules are given in ($\#\mathrm{Outer~Iterations}$, [$\#\mathrm{Inner~Iterations}$]). 
 }
	\label{fig:ber_coded}
 \vspace{-0.6cm}
\end{figure}

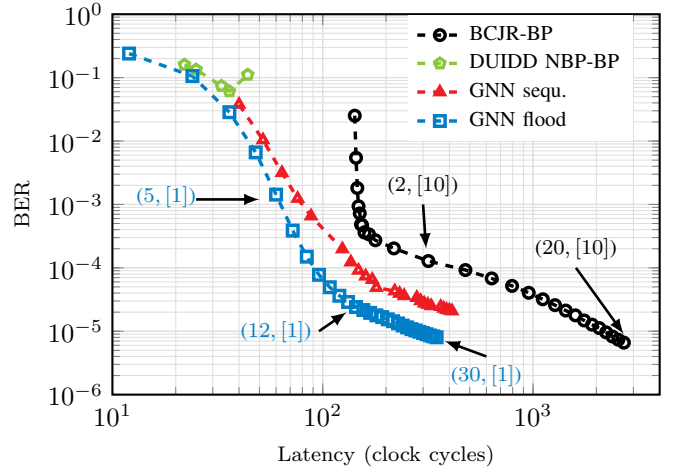
\begin{figure}[t]
	\centering
	\begin{tikzpicture}
		
		\pgfplotsset{compat=1.5}

		\begin{axis}[
			xmode=log,
			ymode=log,
			xlabel=\footnotesize $\mathrm{ Latency~(clock~cycles)}$, %
			ylabel=\footnotesize $\mathrm{BER}$,
			xmin = 10,
			xmax=4*10^(3),
			ymax=10^(0),
			ymin=10^(-6),
			mark size=2.5pt,
			legend style={at={(axis cs:3000,1.0)},anchor=north east, draw=none,
            fill=white,
            fill opacity=0.7,}, 
			grid=both,
			minor grid style={gray!25},
			major grid style={gray!25},
			width=\linewidth,
	        height=0.75\linewidth,
			cycle list name=corporate colours markers,
			legend cell align={left},
			line width=0.8pt, %
			]

            \addplot+ [black, dashed,mark options={solid},mark=o, line width=1.3pt, mark size= 2pt] 
			table[x expr=\thisrowno{0}+0.00 ,y=ber,col sep=comma]{./tikz/results/latency/latency_bcjr_bp_10_idd_20.txt};
			\label{plot:latency_bcjr_bp_big_iters_bp_10_outer_iters_20}
            \addlegendentry{\footnotesize BCJR-BP};
            
            \addplot+ [apfelgruen, dashed,mark options={solid},mark=pentagon, line width=1.3pt, mark size= 2pt] 
			table[x expr=\thisrowno{0}+0.00 ,y=ber,col sep=comma]{./tikz/results/latency/latency_duidd_3_bp_nn_3_spa_8.txt};
			\label{plot:latency_duidd_3_bp_nn_3_sp_8_snr_15_17_new}
            \addlegendentry{\footnotesize DUIDD NBP-BP};

            \addplot+ [rot, dashed,mark options={solid},mark=triangle, line width=1.3pt, mark size= 2pt] 
			table[x expr=\thisrowno{0}+0.00 ,y=ber,col sep=comma]{./tikz/results/latency/latency_gnn_seq.txt};
			\label{plot:latency_gnn_seq}
            \addlegendentry{\footnotesize GNN sequ.};
            
            \addplot+ [mittelblau, dashed,mark options={solid},mark=square, line width=1.3pt, mark size= 2pt] 
			table[x expr=\thisrowno{0}+0.00 ,y=ber,col sep=comma]{./tikz/results/latency/latency_gnn_flooding.txt};
			\label{plot:latency_gnn_flooding}
            \addlegendentry{\footnotesize GNN flood};

            \node[] at (axis cs: 18, 1.3e-3) {\footnotesize \textcolor{mittelblau}{($5,[1]$)}};
            \draw[-latex, line width=1pt, color=black] (axis cs: 25, 1.2e-3) --node[below,yshift=0cm, xshift=.1cm, color=black]  {} (axis cs: 50, 1.2e-3);
            
            \node[] at (axis cs: 60, 1e-5) {\footnotesize \textcolor{mittelblau}{($12,[1]$)}};
            \draw[-latex, line width=1pt, color=black] (axis cs: 96, 1e-5) --node[below,yshift=0cm, xshift=.1cm, color=black]  {} (axis cs: 130, 2e-5);

            \node[] at (axis cs: 600, 2e-6) {\footnotesize \textcolor{mittelblau}{($30,[1]$)}};
            \draw[-latex, line width=1pt, color=black] (axis cs: 600, 4e-6) --node[below,yshift=0cm, xshift=.1cm, color=black]  {} (axis cs: 380, 7e-6);

            \node[] at (axis cs: 1600, 2e-4) {\footnotesize \textcolor{black}{($20,[10]$)}};
            \draw[-latex, line width=1pt, color=black] (axis cs: 1600, 1e-4) --node[below,yshift=0cm, xshift=.1cm, color=black]  {} (axis cs: 2700, 9e-6);

            \node[] at (axis cs: 300, 2e-3) {\footnotesize \textcolor{black}{($2,[10]$)}};
            \draw[-latex, line width=1pt, color=black] (axis cs: 300, 1e-3) --node[below,yshift=0cm, xshift=.1cm, color=black]  {} (axis cs: 310, 2e-4);

		\end{axis}

	\end{tikzpicture}
	\caption{\footnotesize \ac{BER} of uncoded bits $\uv$ using coded system  ($N=132,~R=0.5$, $5$G \ac{LDPC},  BPSK) over latency at \ac{SNR} $12\operatorname{dB}$ of different equalizers and decoders over the Proakis-C channel. The \ac{JED} schedules are given in ($\#\mathrm{Outer~Iterations}$, [$\#\mathrm{Inner~Iterations}$]) }
	\label{fig:latency}
 \vspace{-0.6cm}
\end{figure}

For showcasing the potential of \acp{GNN} for \ac{JED}, especially in low latency scenarios, we evaluate a transmission encoded by a 5G \ac{LDPC} code ($K=66, N=132, R=0.5$) and \ac{BPSK} signaling over the Proakis-C channel.
In Fig.~\ref{fig:ber_coded} the \ac{BER} of the bits $\uv$ is shown over the \ac{SNR}. 
For almost all systems, two variants are shown.
The dashed ones indicate systems disregarding latency and focusing on error correcting performance.
The solid ones share a common latency constraint of $144$ clock cycles, equivalent to $12$ iterations of the \ac{GNN} with flooding schedule.
The resulting schedules are given in the legend in the form ($\#\mathrm{Outer~Iterations}$, [$\#\mathrm{Inner~Iterations}$]).
For an equal latency, the flooding schedule \ac{GNN} outperforms the sequential \ac{GNN} and the iterative \ac{BCJR}-\ac{BP} baseline by \qty{2.25}{dB}.
In the all-out performance case, the \ac{BCJR}-\ac{BP} baseline outperforms the flooding \ac{GNN} by \qty{0.15}{dB}, however, the latency is $7.7$ times higher.
Unexpectedly, the flooding \ac{GNN} outperforms the sequential \ac{GNN} regardless of the latency constraints, suggesting that the flooding schedule may be advantageous during training. 
The only \ac{BP} variant with the classical update equations that showed good enough performance to appear on the graph is a variant with a \ac{NBP} equalizer, a \ac{BP} decoder and trainable weights at every edge between the components (called \ac{DUIDD}) \cite{wiesmayr2022duidd}.
Fig.~\ref{fig:latency} displays the \ac{BER} over the latency in clock cycles. 
Recall that  $12$, $2$, and $N+L+2$ clock cycles are needed for the flooding \ac{GNN}, \ac{BP} and \ac{BCJR}, respectively.
In the low latency regime, the flooding \ac{GNN} outperforms the sequential \ac{GNN} and the \ac{BCJR}-BP baseline, as it demonstrates a \ac{BER} $3$ times lower than the sequential \ac{GNN} and $10$ times lower than the \ac{BCJR}-\ac{BP} baseline.
The flooding \ac{GNN} is only outperformed by the \ac{BCJR}-BP baseline after $\sim7$ times the latency of the \ac{GNN}.

\section{Conclusion}
The paper proposes and demonstrates the application of \acp{GNN} for equalization and \ac{JED}. 
In the case of equalization, the \ac{GNN} based on the \ac{FFG} shows almost optimal performance at a fixed latency compared to \ac{BCJR} equalization whose latency depends on the transmission length.
Thus, \acp{GNN} may be an elegant way of bridging the gap between \ac{BP} and \ac{BCJR} for equalization.
In the case of \ac{JED}, we demonstrated the end-to-end learning capabilities of \acp{GNN} for equalization and decoding, leveraging sparse graphs and the blessing of dimensionality. 
With the proposed flooding schedule, the \ac{GNN} outperforms an iterative \ac{BCJR}-\ac{BP} baseline with a substantially lower \ac{BER} for a constrained latency.
Future works may analyze generalization, adaptability, and robustness aspects of the \ac{GNN}-based equalization.
In addition, scalability to larger block lengths, higher modulation orders, and channels with longer memory are interesting aspects.


 \IEEEtriggeratref{16}

\bibliographystyle{IEEEtran}
\bibliography{references}

\end{document}